\documentclass[manuscript,screen]{acmart}

\AtBeginDocument{%
  \providecommand\BibTeX{{%
    \normalfont B\kern-0.5em{\scshape i\kern-0.25em b}\kern-0.8em\TeX}}}

\setcopyright{acmcopyright}
\copyrightyear{2021}
\acmYear{2021}
\acmDOI{TBA}
\usepackage{xcolor}

\usepackage{subfig}

\acmConference[Conference acronym 'XX]{Make sure to enter the correct
  conference title from your rights confirmation emai}{June 03--05,
  2018}{Woodstock, NY}
\acmPrice{15.00}
\acmISBN{978-1-4503-XXXX-X/18/06}

\begin{document}

\title{User Engagement and the Toxicity of Tweets}
\author{Nazanin Salehabadi}
\affiliation{%
  \institution{The University of Texas at Arlington}
  \streetaddress{500 UTA Blvd}
  \city{Arlington}
  \state{Texas}
  \country{USA}
  \postcode{76013}
}
\author{Anne Groggel}
\affiliation{%
  \institution{North Central College}
  \streetaddress{30 N. Brainard Street}
  \city{Naperville}
  \state{Illinois}
  \country{USA}
  \postcode{60540}
}
\author{Mohit Singhal}
\affiliation{%
  \institution{The University of Texas at Arlington}
  \streetaddress{500 UTA Blvd}
  \city{Arlington}
  \state{Texas}
  \country{USA}
  \postcode{76013}
}
\author{Sayak Saha Roy}
\affiliation{%
  \institution{The University of Texas at Arlington}
  \streetaddress{500 UTA Blvd}
  \city{Arlington}
  \state{Texas}
  \country{USA}
  \postcode{76013}
}
\author{Shirin Nilizadeh}
\affiliation{%
  \institution{The University of Texas at Arlington}
  \streetaddress{500 UTA Blvd}
  \city{Arlington}
  \state{Texas}
  \country{USA}
  \postcode{76013}
}

\renewcommand{\shortauthors}{Salehabadi et al.}

\begin{abstract}
Twitter is one of the most popular online micro-blogging and social networking platforms. This platform allows individuals to freely express opinions and interact with others regardless of geographic barriers. However, with the good that online platforms offer, also comes the bad. Twitter and other social networking platforms have created new spaces for incivility. With the growing interest on the consequences of uncivil behavior online, understanding how a toxic comment impacts online interactions is imperative. 
We analyze a random sample of more than 85,300 Twitter \emph{conversations} to examine differences between toxic and non-toxic conversations and the relationship between toxicity and user engagement. We find that toxic conversations, those with at least one toxic tweet, are longer but have fewer individual users contributing to the dialogue compared to the non-toxic conversations. However, within toxic conversations, toxicity is positively associated with more individual Twitter users participating in conversations. This suggests that overall, more visible conversations are more likely to include toxic replies. 

Additionally, we examine the sequencing of toxic tweets and its impact on conversations. Toxic tweets often occur as the main tweet or as the first reply, and lead to greater overall conversation toxicity. 
We also find a relationship between the toxicity of the first reply to a toxic tweet and the toxicity of the conversation, such that 
whether the first reply is toxic or non-toxic sets the stage for the overall toxicity of the conversation, following the idea that hate can beget hate.

\end{abstract}




\maketitle
\section{Introduction}
Social media allows individuals to freely express opinions, engage in interpersonal communications, and learn about new trends and new stories. However, these platforms also create spaces for uncivil behavior. 
Mediated communication restricts social cues, such as body language, that help to discourage negative behavior and instead allows for greater perceived anonymity, which can lead to more toxic behaviors online~\cite{citron2014hate}. 
In particular, toxicity as explicit language, derogatory, or disrespectful content has become endemic on online platforms~\cite{papacharissi2002virtual}. 
This has created a growing concern over incivility on social media platforms~\cite{rost2016digital}. 
Twitter, as one of the leading micro-blogging social media platforms, provides an important opportunity to examine differences in the dynamics of \emph{toxic} and \emph{non-toxic} conversations. 

In this paper, we assess a random sample of \emph{tweets} to examine how social cues provided from toxic main tweets or direct replies may influence subsequent exchanges. For instance, a toxic tweet or direct reply can set the tone of a conversation by serving as a social cue, indicating the permissibility of such language to other users. We highlight the differences between \emph{toxic} and \emph{non-toxic} Twitter conversations as well the \emph{less toxic} and \emph{more toxic} conversations.  
This is accomplished by utilizing Google's Perspective API~\cite{jigsaw2018perspective}, as a state-of-the-art toxicity detection tool, to assign toxicity scores to main tweets and its direct replies. 

Using this tool allowed us to create a binary toxic variable, in which conversational exchanges are grouped into two categories: \emph{toxic} conversations are those with at least one toxic tweet, and \emph{non-toxic} conversations are those with no toxic tweet. In the context of this study, we refer to conversations as the immediate \emph{direct replies} a tweet garners within a 48 hour window. By classifying these conversations or exchanges by toxicity, we compare the length of conversations and the number of unique Twitter accounts participating posting direct replies, while controlling for a number of factors that could affect these measures of engagement, including, the characteristics of the Twitter account that posted the root tweet and the averaged friends, followers, and lists of users directly replying to the root tweet.
We also construct a continuous measure for toxicity called \emph{toxicity score}. We calculated toxicity by taking the total number toxic tweets and dividing it by the total number of tweets (both the main tweet and direct replies). This continuous toxicity measure restricts the data to toxic conversations and allows us to explore the association between conversation length and toxicity. How online communities react to divisive behavior reflects social norms- either spreading negative behaviors, or calling out racist or sexist behavior, or ignoring toxic behaviors~\cite{binns2012don}. 
Therefore, we track the sequencing of toxic tweets to examine how the location (index) of a toxic main tweet or direct reply impact user engagement overall. 
We first examine the relationship between the location of the first initial toxic tweet (either as the main tweet or a toxic direct reply) and user engagement after this initial toxic tweet. We then investigate the relationship between the first \emph{direct reply} to the first toxic tweet and user engagement. With these aims we propose the following five hypotheses in the next section.

\section{Hypotheses}
Both online and offline, individuals are often concerned with their presentation of self and seek to maintain a positive impression from others. 
Online platforms may lessen this concern and may encourage incivility. 
Anonymity provided through online platforms can lead to disinhibition~\cite{suler2004online}, deindividuation, and a lack of accountability~\cite{brown2018so,lee2015people}, which can encourage negative behavior~\cite{lowry2016adults,siegel1986group,kiesler1984social,fichman2019impacts}.
Communicating with others on Twitter may reduce the potential backlash one might experience when making toxic comments in a face-to-face conversation. Even if users' Twitter accounts are linked to their offline identities, such as their name, it is unlikely that someone will meet the person they reply to. By transcending physical interactions, online users can avoid the consequence of face-to-face confrontation. 
Moreover, the nature of online platforms with sending instant replies make online platforms more apt to capture unfiltered prejudices~\cite{brown2018so}. The expectations of behavior for conversations and the consequences when individuals breach those norms, that would keep toxic behavior in check offline, often do not apply online.

Hansen et al.~\cite{hansen2011good} found that, on Twitter, news are retweeted more if they portray some negative content or opinion, while negative sentiment was detrimental to retweeting in a generic setting. 
Toxic conversations might gain visibility in a similar way with more individuals joining on. 
Or, given the triggering nature of toxic tweets, they might lead to longer conversations as more Twitter users counter that language and speak out. 
Or, as in the case of social media trolls, users may opt to ignore negative
behavior~\cite{binns2012don,shachaf2010beyond,herring2002searching} rather than increasing the visibility of toxic content. 
Given these bodies of research, we propose the following two hypotheses: 
\begin{itemize}

   \item[] \emph{\textbf{H1.} Toxic conversations and non-toxic conversations will have significantly different lengths, i.e., number of tweets.}
    
    \item[] \emph{\textbf{H2.} Toxic conversations and non-toxic conversations will have significantly different levels of user participation.}
\end{itemize}

Twitter users can customize their profiles through adding profile pictures, bio descriptions, and URL links. Such changes can provide cues to users' real identities, as other Twitter users may opt to not make these changes to their profiles. Since online factors such as anonymity or perceived anonymity can foster uncivil behavior~\cite{wulczyn2017ex}, we would expect that Twitter users who are less identifiable are more likely to join toxic conversations. 
Thus, we hypothesize that:
\begin{itemize}
    \item[] \emph{\textbf{H3.} Twitter accounts with less identifiability will be more likely to participate in Toxic conversations than accounts that provide a profile image, URL, have a verified account, or provide other potential markers of their real identity.}
\end{itemize}

However, what extent do other Twitter users join in? 
Other work shows that observing trolling behavior by others influences new users~\cite{cheng2017anyone}. In other words, individuals may be more likely to engage in toxic exchanges after seeing others do it, believing it to be the norm. Does Twitter follow a similar pattern with conversations turning toxic if the initial tweet is toxic? An initial toxic tweet or toxic replies may serve as cues, indicating to Twitter users that toxic content is permissible. Thus, we propose our final two hypotheses:

\begin{itemize}

    \item[] \emph{\textbf{H4.} If the initial tweet is toxic then the conversation is more likely to be toxic.}
    \item[] \emph{\textbf{H5.} If the reply to an initial toxic tweet is toxic then the rest of conversation is more likely to get toxic.}
\end{itemize}
These hypotheses highlight the importance of understanding intergroup dynamics or group mentality in a digital age by investigating the effects of a toxic main tweet or toxic replies. As stated earlier, these direct replies reflect the immediacy of online conversations given that data collection was restricted to a 48 hour window.
Examining these patterns of direct replies outlines potential avenues to counter toxicity. If we understand the nature of immediate direct replies, greater efforts can be made to address toxicity on social media platforms.

\section{Related Work}
\textbf{Detection and Classification}. 
Research on toxicity has employed machine learning based detection algorithms to identify and classify offensive language, hate speech, and cyberbully~\cite{zhang2016conversational,davidson2017automated}. For example, Koratana and Hu~\cite{koratanatoxic} classified comments into seven groups of clear, toxic, obscene, insult, identity hate, severe toxic, and threat. 
The machine learning methods use a variety of features, including lexical properties~\cite{nobata2016abusive,mehdad2016characters,de2021multilingual}, demographic and geographic features~\cite{waseem2016hateful,vijayaraghavan2021interpretable}, sentiment scores~\cite{gitari2015lexicon,dinakar2012common,sood2012profanity}, paragraph embeddings~\cite{nobata2016abusive,djuric2015hate}, and linguistic psychological features~\cite{elsherief2018hate}. The state-of-the-art toxicity detection tool is available through, although the granular details of this tool have not been published, broadly speaking, the API which uses machine learning techniques to identify toxicity in text~\cite{jigsaw2018perspective}. 
Another tool that has been used by researchers in detecting offensive language and hate speech is HateSonar~\cite{davidson2017automated}, which is a logistic regression-based classifier~\cite{elsherief2018hate}. 

In this project, we will use this API to detect the toxicity of a sample of main tweets and direct replies.

\textbf{Characterization.} Several approaches have been proposed to measure abusive behavior on social media platforms. Chen et al.~\cite{chen2012detecting} used both textual and structural features to predict a user’s intrinsic desire in producing toxic content in YouTube, while others~\cite{kayes2015social} found that users tend to flag toxic content posted on Yahoo Answers in an exorbitance correct way. Also, some users considerably deviate from community norms, posting a large amount of toxic content. Through careful feature extraction, 
They also used machine learning approaches to predict which users will be suspended. 
Chatzakou et al.~\cite{chatzakou2017mean} studied the properties of bullies and aggressors, focusing on the users of tweets with the \#GamerGate hashtag.~\cite{elsherief2018peer} compared the characteristics of hate instigators and hate targets and showed both hate instigator and target users are more likely to get attention on Twitter, i.e., get more followers, retweets and listed.
Others have studied the prevalence and characteristics of hate speech on specific web communities, such as Gab~\cite{zannettou2018origins}, 
4chan's Politically Incorrect board (/pol/)~\cite{hine2016kek}, Twitter~\cite{elsherief2018hate,chatzakou2017mean,finkelstein2018quantitative,singhal2022sok} and Whisper~\cite{silva2016analyzing}.

\textbf{Impact.} A few studies have examined the impact of hate speech on hate targets and the impact of toxicity on flow of conversations. 
For example, some work have examined the association between hate speech and contributing factors, including terrorist attacks~\cite{olteanu2018effect,williams2015cyberhate}, crime~\cite{wired2016insideGoogle}, 
and political events~\cite{hine2017kek,harlow2015story}. For example,~\cite{kwak2015exploring} shows that online games make players particularly vulnerable to the exhibition of, and negative effects from, cyberbullying and toxic behavior. ~\cite{zhang2018conversations} examined conversation dynamics in an Oxford-style debate dataset. 
Their findings showed the outcome of debate depends on aspects of conversational flow, including number of discussion points, talking points, and discussion feedback. 
Maity et al.~\cite{maity2018opinion} studied the factors associated with incivility on Twitter, which caused users to leave Twitter. They found that the act of incivility is highly correlated with the opinion differences between the account holder (i.e., the user writing the incivil tweet) 
and the target (i.e., the user for whom the incivil tweet is meant for or targeted),

\textbf{Dialogues in Social Media.} 
Dialogue modeling has been extensively studied by researchers. 
~\cite{kim2014towards} presents an approach for classifying student discussions according to a set of discourse structures, and identifying discussions with confusion or unanswered questions.
Works such as~\cite{kittur2008harnessing,medelyan2009mining,stvilia2008information,ferschke2012behind,panciera2009wikipedians} have studied dialogue modeling in Wikipedia talk pages. 
Previous works have also looked at how MOOC discussion forums can be modeled to categorize threads based on whether or not they are  related to course content~\cite{d2012dynamics,chaturvedi2014predicting,brinton2014learning,wise2017mining,agrawal2015youedu,kumar2015learning,chaturvedi2014predicting,rossi2014language,stump2013development}.
Other works have proposed some approaches for modeling dialogue acts in conversational speech~\cite{stolcke2000dialogue,artstein2008inter,rojasbarahona-EtAl:2019:W19-59,mairesse2007using, liu2006enriching,rose2008analyzing,platonov-EtAl:2020:sigdial},
or have classified comments into a set of coarse discourse acts to better understand the conversation~\cite{zhang2018making,jain-EtAl:2020:sigdial,chang-EtAl:2020:sigdial,tan2016winning, cheng2017factored,zhang2017characterizing,kim2010tagging,michael:2020:sigdial}. 
Some studies also have focused on emotion recognition in conversations~\cite{wang-EtAl:2020:sigdial1,cercascurry-rieser:2019:W19-59}. 
Scholars have explored detecting toxicity of conversations~\cite{zhang2018characterizing,jurgens2019just,aroyo2019crowdsourcing,chang2019trouble,zhang2018conversations,hessel2019something}. %
Despite this scholarship, understanding the dynamics of toxic main tweets or toxic direct replies and subsequent exchanges has been understudied.

\section{Data Collection}
We used Twitter Standard API to collect a random sample of public tweets during the period of April 24th to 26th 2020. In total, we obtained more than 3.6 million tweets. From this sample, we removed retweets and kept the \emph{quote tweets}, which are similar to retweets except that they include a new tweet message. 
We removed retweets from the data to focus on conversational content with main tweets and direct replies. We restricted our sample of \emph{main} tweets to those in English because the Perspective API can determine the toxicity of text for a limited set of languages~\cite{fernquist2019study,parekh2017toxic,noever2018machine}. 

This can increase the probability of obtaining mainly direct replies in English. We obtained an initial sample of 441,152 tweets. Then, we employed the \emph{replies} function included in the Twarc library~\cite{twarc}, a Python library for obtaining Twitter data. This tool allowed us to obtain the direct replies within 48 hours after the main tweet was posted in our sample. 
We collected the direct replies to the tweets in our sample, which we call ``main'' or ``root'' tweets.   
Twitter Standard search API returns all tweet replies for a particular tweet up to a 7 day timeframe from the first day of the data collection.
While some tweets receive countless replies as soon as they are posted, others have a slower rate of replies. Therefore, it is important to note that: First, we collected the replies after a few days, so that main tweets have enough time to get responses. Second, we provided conversations the same amount of time to receive replies. As noted previously, our dataset is restricted to direct replies posted in the first 48 hours after the main tweet was posted. Tweets that did not have a reply, or conversations that were generated by a single user were removed. 
Thus, our final dataset includes 85,389 conversations and 257,694 tweets, where the conversation length ranges from 2 to 6,101 tweets.

\subsection{Conversation Toxicity Detection} 
To identify toxic main tweet and toxic direct replies in our data, we ran tweets through the Google Perspective API~\cite{jigsaw2018perspective}, which has been used by many scholars~\cite{zannettou2020measuring,grondahl2018all,elsherief2018hate}. Google perspective API applies different machine learning models to score the toxicity of textual data, including \emph{Toxicity}, and \emph{severe-toxicity}. 
\emph{Toxicity} is defined as rude, disrespectful, or unreasonable comment and \emph{severe-toxicity} is defined as a very hateful, aggressive, disrespectful comment. 
We randomly selected 200 tweets and used both models for detecting their toxicity scores. 
Two coders then separately classified the tweets as \emph{toxic} and \emph{non-toxic}. For inconsistent results, coders discussed how to resolve disagreements. The inter-rater agreement was 94\% with a Cohen's kappa of 0.61. Using this manual labeling, we computed the accuracy of \emph{toxicity}, and \emph{severe-toxicity} models, which are 94\% and 96\%, respectively.  
The inter-rater reliability showed substantial agreement between manual coders and the Perspective API classification for for both \emph{toxicity} and \emph{severe-toxicity} models. Specifically, results suggest substantial agreement with a Cohen’s kappa of 0.64 for toxic tweets and 0.69 for severe toxicity. 
Since the accuracy and the agreement between the \emph{severe-toxicity} scores of the Perspective API and the coders was higher, we used this model to detect toxicity score for every tweet in the conversations.  

Before processing the tweets through the Google Perspective API, we pre-processed the tweets by removing URLs and punctuation. However, we did not remove hashtag keywords as they might contain toxic content. Additionally, emojis were included in our sample because they may show Twitter users' emotions, which can contribute to the toxicity of the tweets~\cite{riordan2017emojis}. 
We converted emoticons or emojis into text using a Python module called demoji~\cite{emoji}. Although the main tweets in our sample were restricted to English, since the sequencing of replies was an area of interest, therefore direct replies were included regardless of language. 
This was accomplished by using Google Translate API~\cite{googletrans} to translate tweets written in other languages into English, and then the translated text was passed through the Perspective API to get their sever toxicity scores. 
We identified 40,007 tweets in other languages, i.e., about 15\% of all the tweets in our dataset. 

\subsection{Data Modeling} 
Although we were unable to collect nested conversations due to Twitter API limitations, examining nested conversations might reflect a different pattern of interactions than the ones examined in our hypotheses. For instance, replies to a reply can be viewed as new conversations since what motivates users to engage in those sub-conversation may differ from what prompts users to engage in the main conversation. While it is likely that Twitter exchanges continue and evolve under replies, that is beyond the scope of this study. Limiting direct replies within a 48 hour window allowed us to better examine effects of the toxicity of the main tweet on immediate replies. For instance, individuals may be more likely to post toxic replies after seeing a toxic main tweet or after reading toxic direct replies. If individuals post toxic tweets because they perceive it to be acceptable within a conversation, those perceptions may perhaps be strongest when exchanges reflect temporal immediacy. 

Since in our hypotheses, the order of the direct replies plays a role, we define a \emph{Twitter Conversation} as a time series variable, $C=(t, r_1, r_2, ...)$, where $t$ is the main (root) tweet in the conversation, $r_1$ is the first reply (or, response) to this tweet, $r_2$ is the second direct reply to this tweet, and so on. The length of conversation specifies the number of tweets in the conversation and can be two or more, $len|C|\geq 2$. 

\subsection{Dataset Characterization} 
\begin{figure}[t!]
\center
\subfloat[CCDF of convs. length]{\includegraphics[width=0.28\columnwidth]{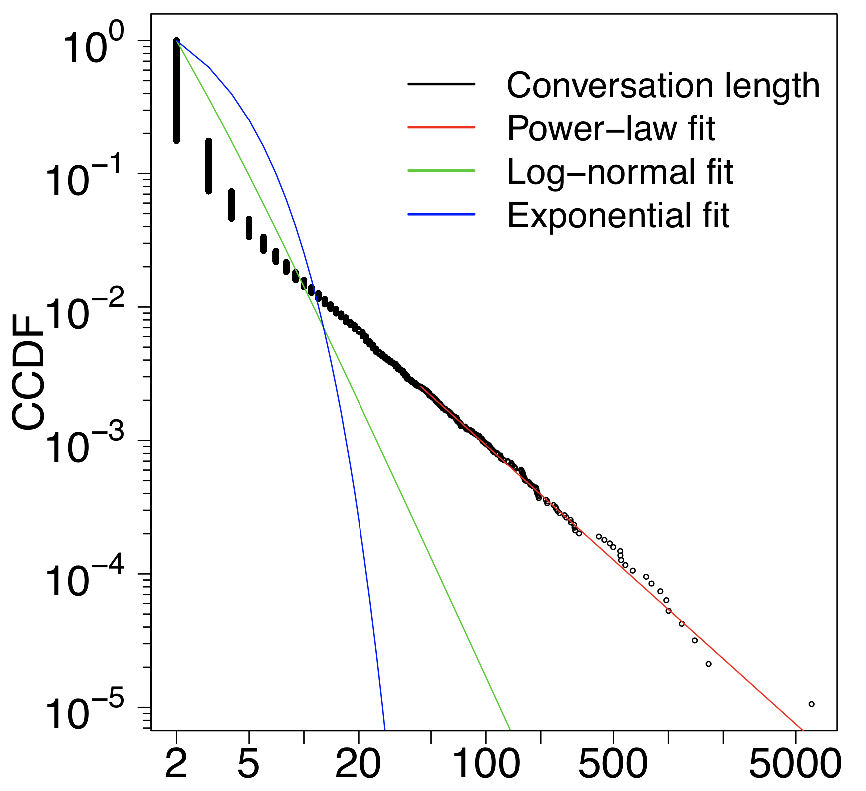}\label{fig:2}}
\hspace{2mm}
\subfloat[Toxicity score]{\includegraphics[width=0.28\columnwidth]{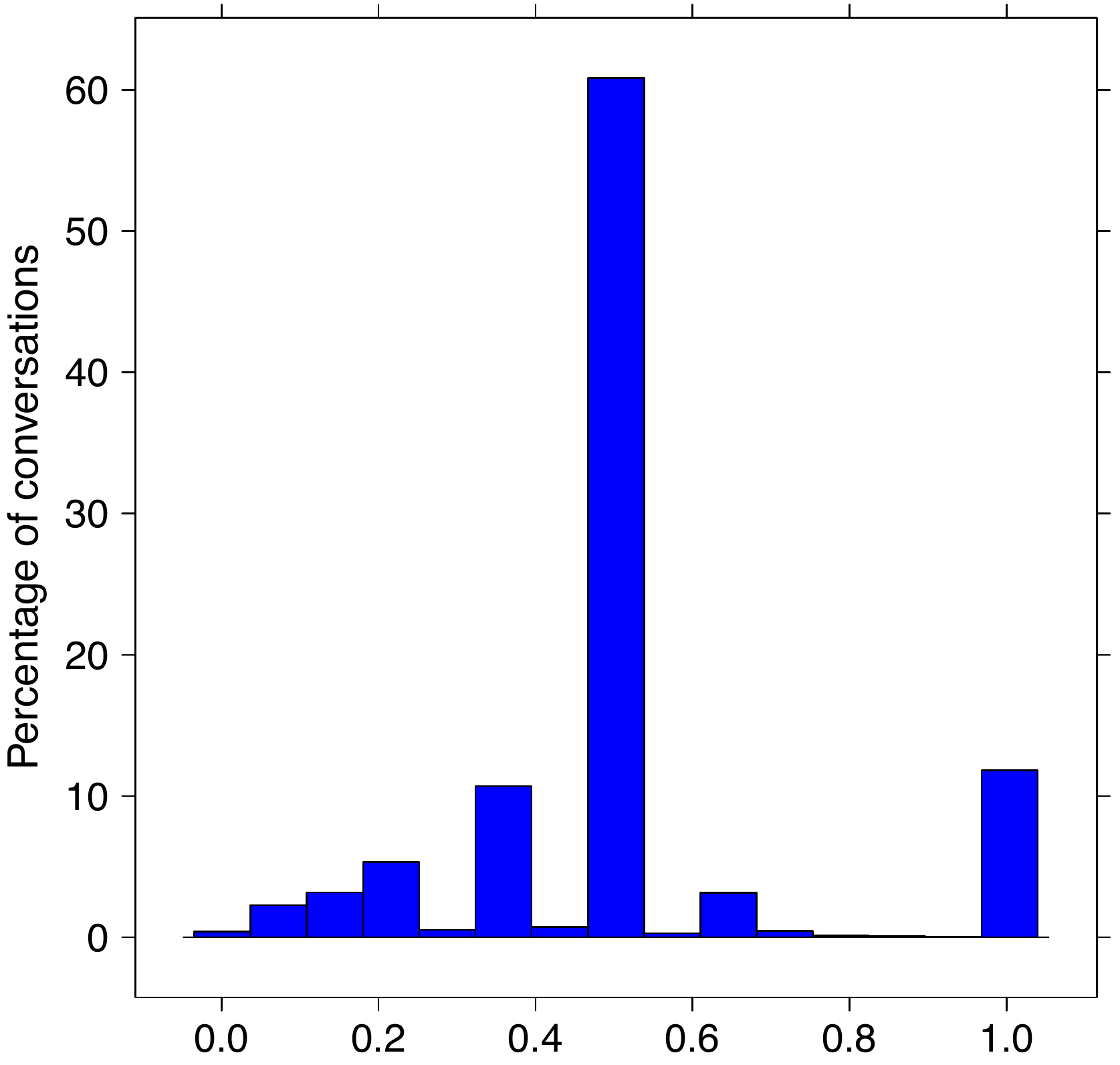}\label{fig:96}} 
\caption{Dataset Characteristics}
\vspace{-4mm}
\end{figure}
Our final dataset includes 85,389 conversations with 257,694 tweets, posted by 227,658 unique Twitter users. 
In our dataset, conversation length ranges from 2 to 6,101 tweets, while the majority of conversations (about 81\%) have a length of 2. 
Figure~\ref{fig:2} shows the histogram of length of conversations. 
CCDF plot for conversation length in our dataset, where it is fitted to a discrete power law, with $x_{min}= 9$ and $\alpha = 2.21$. 
To test if the power-law distribution is a good fit to our data, we test the power-law hypothesis using goodness-of-fit test~\cite{clauset2009power}. 
We use Kolmogorov-Smirnov statistic to measure the distance between the distribution of the conversation length and 5,000 synthetic datasets from a true power-law distribution. 
The $p$-value quantifies the plausibility of the hypothesis which is defined to be the fraction of the synthetic distances that are larger than the empirical distance. If $p$ is close to 1, then the difference between the empirical data and the model can be attributed to statistical fluctuations alone; if it is small, the model is not a plausible fit to the data. 
We obtained $p=0.40$, which indicates that the power-law distribution is a plausible distribution for the length of conversations. 
Even though our data is well fit by a power law, 
another distribution, such as a log-normal, might give a fit as good or better. Therefore, we compared two models by employing Vuong’s test, which is a likelihood ratio test for model selection using the Kullback-Leibler criteria. 
Our null hypothesis is that both distributions are equally far from the true distribution. Running the test, we could not reject the null hypothesis ($p\_two\_sided =0.89$, $p\_one\_sided =0.55$), which means one model is not closer to the true distribution than the other. Therefore we cannot say power law distribution is a better model for the distribution of conversation length. 
We also collected metadata of users who post the main tweets and replied to the main tweets.  Table~\ref{table:1} shows the descriptive statistics of the Twitter users, including their number of followers, friends, tweets, the age of the account (in years), and if they are verified accounts. 
About 1.2\% percent of our sample are verified Twitter accounts. 
Since the number of followers, friends and tweets exhibit heavy-tailed distributions, we report the median as well as the mean. 

\begin{table}[h!]
\caption{Descriptive statistics of users in our conversations.}
\resizebox{0.60\columnwidth}{!}{%
\begin{tabular}{ l| l l l l l}
\hline \hline
 Feature & Type & Min & Max  & Mean & Median\\
 \hline 
Followers & Count & 0 & 46,441,405 & 6,384 & 402\\  
Friends & Count &  0 & 608,163 & 1,190 & 444\\ 
\#tweets & Count  & 1 & 5,030,869 & 21,721 & 6,300\\
Listed Count & Count &0 &206,848 & 37.56 &2\\
Profile Descr. (length) & Count & 0&205 & 69.68 &61\\
Age (Years) & Count & 0 & 14 & 4.71 &4\\
URL& Boolean&0 &1 &0.30 & 0 \\
Profile Image & Boolean &0 &1 &0.34 &NA\\
Verified & Boolean & 0 & 1 &  2,845 (0.012) & NA\\
  \hline  \hline
 \multicolumn{3}{l}{N= 227,658 Unique Twitter accounts}  & & &\\

\end{tabular}
}
\label{table:1}
\end{table}

\section{Analysis} 
Online communities create social norms that can either prescribe civil exchanges or negative behaviors when users believe or desire toxicity to be the norm.
To understand the construction of these norms we compare \emph{toxic} and \emph{non-toxic} Twitter conversations. 
We accomplish this by using Google Perspective's \emph{severe-toxicity} model, to classify 29,458 (11\%) tweets as \emph{toxic} and 228,236 (89\%) tweets as \emph{non-toxic}. 
We find that 65,687 (about 77\%) of conversations do not include any toxic tweet, and 19,702 (about 23\%) conversations include at least one toxic tweet. 
Next, we empirically test our proposed hypotheses about the differences between \emph{toxic} and \emph{non-toxic} conversations. 

We constructed two variables to measure toxicity.First, we use a binary variable of toxicity. Conversations that include a toxic main tweet or a toxic direct reply are 
classified as \emph{toxic}, and those that include no toxic tweets are classified as \emph{non-toxic}.

Second, we construct a continuous variable for toxicity, representing the overall toxicity score of \emph{toxic} conversations. 
Figure~\ref{fig:96} shows the histogram of toxicity scores of \emph{toxic} conversations (percentage), where the toxicity score is computed by dividing the number of toxic tweets in the conversation by the length of conversation. We define the length of conversation as the number of direct replies + 1 (for the main tweet). This figure shows that the majority of \emph{toxic} conversations (more than 60\%) have a toxicity score of 0.5, which means half of the tweets in these conversations are toxic. Also, some conversations (about 12\%) only include toxic tweets. 

\subsection{Toxicity and Number of Direct Replies}

In this study we examine the relationship between the toxicity of conversations and the length of conversations. Specifically, we hypothesize that \emph{toxic} conversations and \emph{non-toxic} conversations have significantly different lengths in terms of the number of tweets. 
\begin{table*}[t]
\centering
\caption{Toxic vs. Non-toxic Conversations}
\label{convs}
\resizebox{0.8\textwidth}{!}{%
\begin{tabular}{lccccc|ccccc}
\hline \hline \\[-2.1ex] 
& \multicolumn{5}{c}{Toxic Conversations} & \multicolumn{5}{c}{Non-toxic Conversations} \\
\hline  \\[-1.8ex] 
 & Count & Min & Mean & Median & Max & Count& Min & Mean & Median &  Max \\  
\hline \\[-1.8ex] 
Length & 19,702  & 2 & 5.19 & 2 & 6,101 & 65,687 & 2 & 2.37 & 2 & 96 \\
Unique users & 95,130  & 2 & 5.02 & 2 & 6,019 & 140,187 & 2 & 2.31 &  2 &  96\\
\hline \hline
\end{tabular}
}
\end{table*}
Table~\ref{convs} shows the length of conversations, where \emph{toxic} and \emph{non-toxic} conversations have a mean length of 5.19 and 2.37, and max length of 6,101 and 96, respectively. 
Since the distribution of conversation length is not normal and it has a long tail, we ran a non-parametric Mann-Whitney U test to examine whether there are significant differences in the length of conversation by toxic and non-toxic conversations. The Mann-Whitney results suggest that, on average, toxic conversations are longer than non-toxic conversations, ($\mu=5.19$, vs. $\mu=2.37$) ($W =741760000$, $p<0.001$).  

Next, we use linear regression to examine the association between length and toxicity of conversations, where the \textit{dependent} variable is conversation length, and toxicity, as a binary variable, is the \textit{independent} variable. 
In our regression model we incorporate the following control variables: the averaged follower count, friend count, and listed count of users participating in the conversation. We also control for if the root user has a verified account, account age (years) of the root Twitter account, if the root account has used a profile image, and the root user's profile description length. 

Table~\ref{linear-toxic-length} shows the results of our linear regression model for length of conversations and the conversation toxicity (as a binary variable).
All control variables were included in the model. 
For toxic conversations, the predicted length of Twitter conversations is approximately 2.7 tweets longer than non-toxic conversations ($p<0.001$).  
Thus, more than establishing that toxic conversations and non-toxic conversations will have significantly different lengths, we establish conversation toxicity as a significant predictor for conversation length. This may be attributed to longer conversations provided greater opportunity for toxic tweets to be included in the conversation. Or toxic conversations, defined as those including those tweets may draw voices of dissent into Twitter conversations as individuals disagree with one another. 

\begin{table}[h] 
\begin{minipage}{0.49\columnwidth}
\centering 
\caption{Relationship between toxicity as a binary variable \\ (toxic vs. non-toxic) and conversation length} 
\label{linear-toxic-length}
    \resizebox{0.90\columnwidth}{!}{%
    \begin{tabular}{@{\extracolsep{5pt}}lll} 
    \\[-1.8ex]\hline 
    \hline \\[-1.8ex] 
    \multicolumn{3}{c}{\textit{Dependent variable: Conversation length}} \\\cline{2-2}
    & Linear Regression & std err \\ 
\hline \\[-1.8ex] 
 Toxic (binary) & 2.68$^{***}$ &  (0.202) \\ 
  Follower count & 0.00001$^{***}$  & (0.00000) \\ 
  Friend count & 0.0001$^{***}$  & (0.00001) \\ 
  Listed count & $-$0.001$^{***}$  & (0.0001) \\ 
  Avg. Followers count & $-$0.00002$^{***}$  & (0.00000) \\ 
  Avg. Listed count & 0.0002  & (0.001) \\ 
  Verified account & 8.931$^{***}$  & (0.619) \\ 
  Account age (years) & $-$0.007  & (0.028) \\ 
  Has profile image & 0.228  & (0.212) \\ 
  Description length (Chars) & 0.004$^{**}$ & (0.002) \\ 
 \hline \\[-1.8ex] 
Observations & 85,389 \\ 
Log Likelihood & $-$395,200.000 \\ 
Akaike Inf. Crit. & 790,422.000 \\ 
\hline 
\hline \\[-1.8ex] 
\textit{Note:}  & \multicolumn{1}{r}{$^{**}$p$<$0.05; $^{***}$p$<$0.001} \\ 
\end{tabular} 
}
\end{minipage}\hfill
\begin{minipage}{0.49\columnwidth}
\caption{The relationship between conversation length and\\ toxicity, as a continuous variable, in \emph{toxic} conversations} 
 \label{lg-toxicity-length} 
\resizebox{.89\columnwidth}{!}{%
\begin{tabular}{@{\extracolsep{5pt}}lll} 
\\[-1.8ex]\hline 
\hline \\[-1.8ex] 
\multicolumn{3}{c}{\emph{Dependent variable: Conversation length}}  \\ 
\cline{2-2} 
 & Linear Regression & std err\\ 
\hline \\[-1.8ex] 
 Toxicity & $-$19.74$^{***}$ & (1.7)\\ 
  Follower count & 0.00001$^{***}$   & (0.00000) \\  
  Friend count & 0.0002$^{***}$  & (0.00004) \\  
  Listed count & $-$0.002$^{***}$ & (0.0003) \\ 
  Avg. Followers count & $-$0.0001$^{***}$& (0.00001) \\  
  Avg. Listed count & 0.001   & (0.003) \\ 
  Verified account & 28.376$^{***}$  & (2.655) \\
  Account age (years) & $-$0.022  & (0.120) \\ 
  Has profile image & 0.609 & (0.920) \\ 
  Description length (Chars) & 0.011 & (0.007) \\ 
 \hline \\[-1.8ex] 
Observations & 19,702 \\ 
Log Likelihood & $-$105,438.100 \\ 
Akaike Inf. Crit. & 210,898.300 \\ 
\hline 
\hline \\[-1.8ex] 
\textit{Note:}  & \multicolumn{1}{r}{$^{***}$p$<$0.001} \\ 
\end{tabular} 
}
\end{minipage}
\end{table} 

However, since conversation length exhibits a skewed distribution, examining the whole population may not capture nuanced patterns within the data. To address this we adopt two approaches: First, we follow existing work that used quantile regression to analyze highly skewed  variables~\cite{yu2003quantile}. For conversation length, we divided our dataset into quartiles and ran multivariate regression models on each quartile. The results in the second and forth quartiles are consistent with our results on the full data, while the results in the other quartiles did not reach a level of statistical significance. Second, we transform conversation length using a log function, and then run multivariate regression models on the whole dataset. This approach yielded the same results, where conversation toxicity was positively associated with conversation length. 
Additionally, to examine whether conversations with a length of 2 impacted our results, we dropped these conversations and re-ran the analyses. Our significant results remain the same. 

\textbf{\emph{Toxicity in toxic conversations.}} 
As shown in Figure~\ref{fig:96}, some Twitter conversations are more toxic than others. 
For example, 2,331 conversations have a toxicity score of 1, while others are much less toxic, with 4,574 conversations have a toxicity score of less than 0.5. 
To better understand the relationship between conversation length and toxicity, we ran additional analyses restricting the data to only toxic conversations. In particular, 
we test the association between toxicity and length in \emph{toxic} conversations using linear regression, where the dependent variable is length, and the independent variable is toxicity as a \emph{continuous} variable. In this model, we also control for the characteristics of the root Twitter account, as well as the average follower and friend counts of all users participating in the conversation. Table~\ref{lg-toxicity-length} displays that toxicity becomes negatively associated with conversation length and for every unit increase in toxicity, a 19 unit decrease in conversation length is predicted, holding all other variables constant ($p<0.001$). This suggests that across toxic conversations, the more toxic conversations become, the fewer direct replies they receive.

We also investigate the impact of a skewed distribution of conversation length on the analysis by running the regression models on each of the four quartiles. Our results show toxicity is negatively associated with length of conversation in the third and fourth quartiles. However, results in the other quartiles were not significant. By transforming conversation length using a log function, we again find that toxicity was negatively associated with conversation. These results remain the same after dropping conversations with a length of 2. 

In summary, we find that toxic conversations are positively associated with length of conversation. However, when analyses were restricted to only toxic conversations, the level of conversation toxicity was negatively associated with the total number of tweets within a conversation. While these results first appear contradictory with our previous results, they make sense intuitively. Toxicity may be a predictor of length as Twitter users feel compelled to join in a toxic conversation, whether their tweet be one of agreement or dissent. Or rather, longer conversations which involve more direct replies have a greater chance of including toxic content. But the nature of toxicity is more nuanced within conversations. The level of toxicity across all conversations classified as toxic suggests that there is a tipping point in conversations wherein Twitter users may be discouraged from joining in the conversation. 

\subsection{Toxicity and User Participation} 
\emph{\textbf{User account characteristics.}} 
We examine the relationship between users' account characteristics (and potential identifiability) and conversation toxicity.
Table~\ref{table:users-stats} displays the descriptive statistics of Twitter account characteristics for toxic and non-toxic conversations.
Table~\ref{toxic-account-characteristics} displays our logistic regression results, where toxicity (binary) is outcome variable, and users' account characteristics is the predictor, while controlling for Twitter users' account age, follower, friends, tweets, and listed counts. 
Table~\ref{toxic-account-characteristics} shows that verified accounts, compared to Twitter accounts without a verified status, have a 0.32 decrease in the log-odds of being in a toxic conversation, holding all other variables in the model constant ($p<0.001$). We also find that accounts that provide a profile image have a 0.26 decrease in the log-odds of being in a toxic conversation, holding other variables constant ($p<0.001$). Twitter accounts that provide a URL have a 0.4 decrease in the log-odds of participating in toxic conversations ($p<0.001$). Likewise, the length of descriptions in profiles is negatively associated with participating in a toxic conversation. 
Thus, we find support for \emph{Hypothesis 3} that: Twitter accounts with less identifiability will be more likely to participate in toxic conversations than accounts that provide a profile image, a URL, have a verified account status, or provide other potential markers of their real identity. 

\begin{table*}
\caption{Descriptive statistics of Users in Toxic vs. Non-toxic Conversations.}
\centering 
\resizebox{0.95\textwidth}{!}{%
\begin{tabular}{lccccc|ccccc}
\hline \\[-1.8ex] 
& \multicolumn{5}{c}{With \textit{no} toxic tweet} & \multicolumn{5}{c}{With more than \textit{{$\geq$ 1}} toxic tweet} \\
\hline \hline \\[-1.8ex] 
Statistic & \multicolumn{1}{c}{Count} & \multicolumn{1}{c}{Min} & \multicolumn{1}{c}{Max} & \multicolumn{1}{c}{Mean}&\multicolumn{1}{c}{Median} & \multicolumn{1}{c}{Count}& \multicolumn{1}{c}{Min} & \multicolumn{1}{c}{Max} &\multicolumn{1}{c}{Mean} &\multicolumn{1}{c}{Median}  \\  
\hline \\[-1.8ex] 
Followers & 140,187 &0 & 30,502,080 & 5,840.98&503&95,130  &0 & 46,441,405 &7,459.34 &292  \\ 
Friends & 140,187  &0 & 508,064 & 1,223.53&453 &95,130 &0 & 608,163 &1,195.16 &412\\ 
\#tweets & 140,187 &1 & 5,030,869 & 24,908.6&7919 &95,130 &1 & 5,031,063 & 18,225.80&4,751 \\ 
Listed Count& 140,187 &0 &123,899 &41.93  & 3&95,130&0 &206,848 &32.23 &1 \\
Profile Description& 140,187 &0 &205 &72.00  & 64& 95,130&0 &193 &67.02 &57 \\ 
Age (years) & 140,187 &0 &14  &4.89&4  &95,130 &0&14& 4.46 &4 \\
URL& 140,187 &0 &1 & 0.34  &NA&95,130&0 &1 &0.24 &NA \\ 
Profile Image&140,187  &0 &1 &0.37  &NA&95,130& 0&1 & 0.28&NA \\ 
Verified& 2,219 (1.58\%) & NA & NA & NA & NA &726 (0.76\%) & NA & NA& NA & NA\\
\hline  \hline 
\end{tabular} 
}
 \label{table:users-stats} 
\end{table*}

\begin{table} \centering 
  \caption{User account characteristics} 
  \label{toxic-account-characteristics} 
\resizebox{0.50\columnwidth}{!}{%
\begin{tabular}{@{\extracolsep{5pt}}lll} 
\hline 
\hline \\[-1.8ex] 
\multicolumn{3}{c}{\textit{Dependent variable: Toxic (binary)}} \\ 
\cline{2-2} 
 & Logistic Regression & Std err\\ 
\hline \\[-1.8ex] 
  Verified account & $-$0.32$^{***}$ & (0.04) \\ 
  Has profile image &  $-$0.26$^{***}$ & (0.01) \\ 
  Has a URL & $-$0.40$^{***}$ & (0.01) \\ 
  Description length (Chars) & $-$0.001$^{***}$ & (0.0001) \\ 
 \hline \\[-1.8ex] 
Observations & 257,694 \\ 
Log Likelihood & $-$170,333.600 \\ 
Akaike Inf. Crit. & 340,687.100 \\ 
\hline 
\hline \\[-1.8ex] 
\textit{Note: } & \multicolumn{2}{r}{$^{***}$p$<$0.001} \\ 
\end{tabular} 
}
\end{table} 
 
\emph{\textbf{User Participation.}} We hypothesize that \emph{toxic} conversations and \emph{non-toxic} conversations have significantly different levels of user participation.
Table~\ref{convs} shows that in total 95,130 and 140,187 users have participated in \emph{toxic} and \emph{non-toxic} conversations, respectively. Note that some users have contributed to both toxic and non-toxic conversations, and thus the sum of number of users in each population is more than the number of unique users in our dataset. 
In order to examine the relationship between toxicity and number of users participating in the conversation, we defined \emph{user participation} as a continuous variable that shows the number of unique users divided by the length of conversation. This variable can get a value between 0 and 1, where larger values reflect higher levels of user participation. 
Note that with this transformation, we address the issue of skewed distributions.

Table~\ref{lg-user-participation-toxic} displays the results of a regression model that examines the effects of toxicity of conversations \emph{(toxic and non-toxic)} on user participation. In this model, we control for the account characteristics of the user who posted the main tweet such as their followers, listed count, whether the user had a profile image, the length of the description, and the account age. 
We find that for toxic conversations the predicted user participation level would be about 0.01 lower than for non-toxic conversations ($p<0.001$). 
\begin{table}[h] 
\centering 
\begin{minipage}{0.48\columnwidth}
\caption{User participation in toxic vs. non-toxic conversations.}
 \label{lg-user-participation-toxic} 
\resizebox{0.9\columnwidth}{!}{%
\begin{tabular}{@{\extracolsep{5pt}}lll} 
\hline 
\hline \\[-1.8ex] 
  \multicolumn{3}{c}{\textit{Dependent variable: User Participation}} \\ 
\cline{2-2} 
 & Linear Regression & Std err\\ 
\hline \\[-1.8ex] 
 Toxic (binary) & $-$0.01$^{***}$   & (0.001) \\ 
  Follower count & $-$0.000 & (0.00) \\ 
  Friend count & 0.00  & (0.00) \\
  Listed count & 0.00   & (0.00) \\ 
  Avg. Followers count & 0.00  & (0.00) \\ 
  Avg. Listed count & $-$0.00   & (0.00)\\ 
  Verified account & $-$0.009$^{***}$   & (0.002) \\ 
  Account age (years) & 0.0001   & (0.0001) \\ 
  Has profile image & $-$0.001   & (0.001) \\ 
  Has a URL & $-$0.002$^{***}$   & (0.001) \\ 
  Description length (Chars) & $-$0.00003$^{***}$ & (0.00) \\ 
 \hline \\[-1.8ex] 
Observations & 85,389 \\ 
Log Likelihood & 104,062.400 \\ 
Akaike Inf. Crit. & $-$208,100.700 \\ 
\hline 
\hline \\[-1.8ex] 
\textit{Note:}  & \multicolumn{2}{r}{$^{***}$p$<$0.001} \\ 
\end{tabular} 
 }
\end{minipage}\hfill
\begin{minipage}{0.48\columnwidth}
\caption{User participation in toxic conversations.} 
  \label{lg-toxicity-user-participation} 
\resizebox{0.91\columnwidth}{!}{%
\begin{tabular}{@{\extracolsep{5pt}}lll} 
\\[-1.8ex]\hline 
\hline \\[-1.8ex] 
\multicolumn{3}{c}{\textit{Dependent variable: User participation}} \\ 
\cline{2-2} 
 & Linear Regression & Std err\\ 
\hline \\[-1.8ex] 
 Toxicity & 0.059$^{***}$ & (0.003) \\ 
  Follower count & 0.00  & (0.00) \\ 
  Friend count & 0.00   & (0.00) \\ 
  Listed count & 0.00  & (0.00) \\ 
  Avg. Followers count & 0.00 & (0.00) \\ 
  Avg. Listed count & $-$0.00 & (0.00)\\ 
  Verified account & $-$0.002 & (0.004) \\ 
  Account age (years) & 0.0002 & (0.0002) \\ 
  Has profile image & $-$0.002 & (0.001) \\ 
  Has a URL & 0.0001 & (0.001) \\ 
  Description length (Chars) & $-$0.00001 & (0.00001)\\ 
 \hline \\[-1.8ex] 
Observations & 19,702 \\ 
Log Likelihood & 21,708.720 \\ 
Akaike Inf. Crit. & $-$43,393.430 \\ 
\hline 
\hline \\[-1.8ex] 
\textit{Note:}  & \multicolumn{2}{r}{$^{***}$p$<$0.001} \\ 
\end{tabular} 
}
\end{minipage}
\end{table} 
However, the effect of toxicity changes direction when we consider user engagement solely within toxic conversations. Table~\ref{lg-toxicity-user-participation} shows that for every unit increase in conversation toxicity, about a 0.06 unit increase in user engagement is predicted, holding other variables constant ($p<0.001$). Thus, \emph{hypothesis~2} that: Toxic conversations and non-toxic conversations will have significantly different levels of user engagement is supported. Moreover, we find that toxicity is a significant predictor for higher levels of user engagement when considering toxic conversations. This suggests the greater toxicity levels are associated with greater user engagement. 

One explanation can be that toxic tweets may incite a reaction, good or bad. The most toxic tweets may lead to more users participating in conversations as users disagree or try to counter that toxicity. Another explanation is that tweets with higher levels of toxicity can serve as a social cue indicating to others that such language or rhetoric is permissible, and emboldens others to join in. However, when comparing toxic conversations to non-toxic conversations, we find that toxic conversations have lower levels of user engagement. This suggests that while toxicity appears to incite engagement, non-toxic conversations still involve more Twitter users.   

\subsection{Location of First Toxic Tweet in Conversations}
\begin{figure}
\centering
\subfloat[All toxic conversations] {\includegraphics[width=0.45\columnwidth]{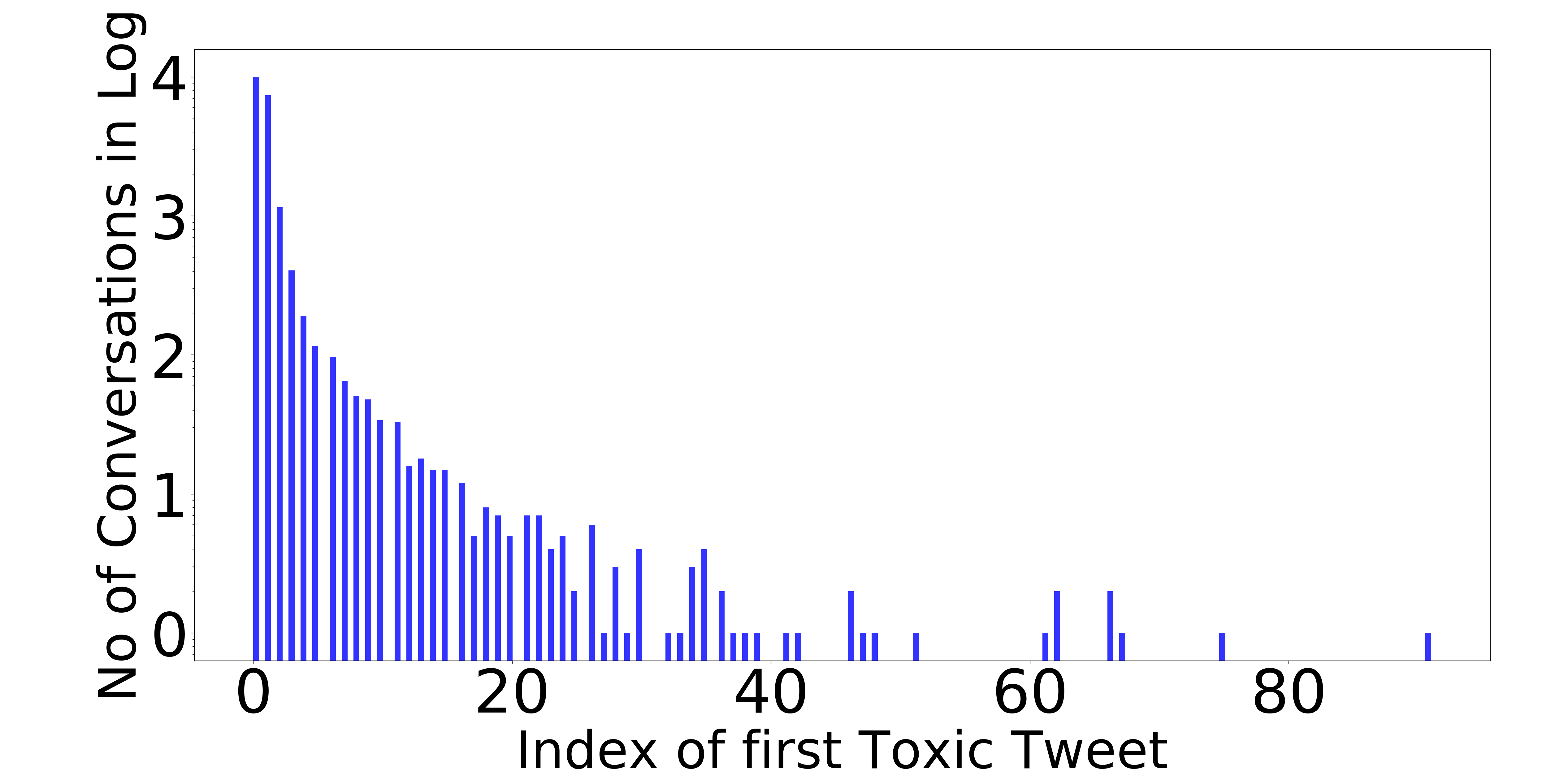} \label{fig:678}}
\subfloat[Excluding length 2 convs.]{\includegraphics[width=0.45\columnwidth]{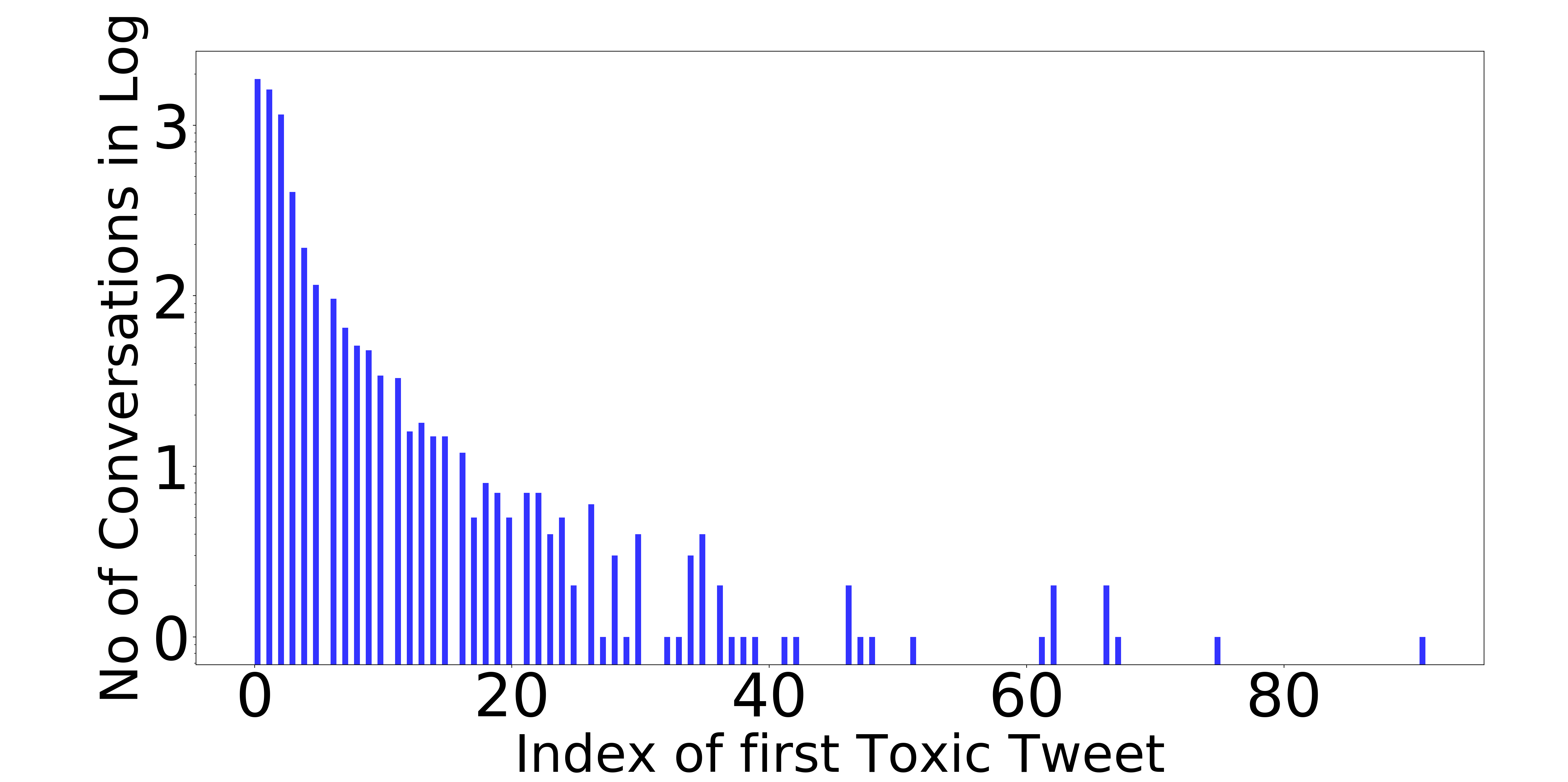} \label{fig:679}} 
\caption{Histogram of index of the first toxic tweet. }
\end{figure}

We examine the location of first toxic tweets in conversations since toxic comments early on may act as social cues, indicating to others that toxic content is permissible. Figure~\ref{fig:678} shows the histogram of index (location) of the first toxic tweet in toxic conversations.
Since about 70\% of toxic conversations have a length of two, we generated the plot after removing conversations with length 2 (Figure~\ref{fig:679}). 
The majority of initial toxic tweets in conversations occur early on. Specifically, within toxic conversations, we find that about 50\% conversations begin with a toxic main tweet. Also, about 50\% of the initial toxic tweets occurred as the first direct reply to the main tweet. 
This suggests that the probability of toxic replies is highest in the initial response to a tweet, even if main tweet is toxic or non-toxic.
Individuals may be more likely to tweet toxic replies earlier on in the conversation because online interactions often lack social cues~\cite{metzger2010social}. 
However, as the conversation evolves, Twitter account users may perceive more conversational cues, and they may be more hesitant to express an unpopular opinion or one that differs from that of the group~\cite{ulbig1999conflict}. 

We hypothesize that if the main tweet is toxic then the conversation is more likely to be \emph{toxic}. We examine the relationship between the toxicity of the main tweet and the overall toxicity of the conversation, only considering toxic conversations. 
Table~\ref{lg-toxicity-first-tweet} shows the results of linear regression, where the conversation toxicity is a continuous variable, and the independent variable is the toxicity of root tweet as a binary variable. In this analysis we include the same control variables as in our previous model. 
We find that toxicity of the main tweet is positively associated with the toxicity of the overall conversation with a coefficient of 0.18 ($p<0.001$). Thus, we find support for Hypothesis~4.  

\begin{table} \centering 
 \caption{Conversation toxicity and first tweet being toxic} 
  \label{lg-toxicity-first-tweet} 
\resizebox{0.5\columnwidth}{!}{%
\begin{tabular}{@{\extracolsep{5pt}}lll} 
\\[-1.8ex]\hline 
\hline \\[-1.8ex] 
\multicolumn{3}{c}{\textit{Dependent variable: Conversation toxicity}} \\ 
\cline{2-2} 
 & Linear Regression & Std err \\ 
\hline \\[-1.8ex] 
 Toxicity of first tweet & 0.18$^{***}$ & (0.003) \\ 
  Follower count & $-$0.00 & (0.00)  \\ 
  Friend count & $-$0.00$^{***}$ & (0.00)  \\ 
  Listed count & $-$0.00  & (0.00) \\ 
  Avg. Followers count & $-$0.00 & (0.00)  \\ 
  Avg. Listed count & 0.00002$^{**}$ & (0.00001) \\ 
  Verified account & $-$0.162$^{***}$ & (0.010) \\ 
  Account age (years) & 0.0001  & (0.0005) \\ 
  Has profile image & $-$0.015$^{***}$ & (0.004)  \\ 
  Has a URL & $-$0.024$^{***}$ & (0.003)  \\ 
  Description length (Chars) & $-$0.0001$^{***}$ & (0.00003)  \\ 
 \hline \\[-1.8ex] 
Observations & 19,702 \\ 
Log Likelihood & 4,468.375 \\ 
Akaike Inf. Crit. & $-$8,912.750 \\ 
\hline 
\hline \\[-1.8ex] 
\textit{Note:}  & \multicolumn{2}{r}{$^{***}$p$<$0.001} \\ 
\end{tabular} 
 }
\end{table}

\subsection{Immediate Response to Toxicity}
Toxic behavior online impacts both the tone and level of interaction.  
Previous work has classified bystander behavior in the face of cyberbullying, as a form of toxicity, into confrontation responses and supportive behaviors~\cite{machavckova2013bystanders}. We seek to examine responses to toxic comments or aggressive behavior on social media platforms. 

We begin by examining how Twitter users respond to a toxic tweet in conversations. Out of the 19,702 toxic conversations, we find that 3,220 (16\%) and 9,779 (50\%) conversations have an immediate \emph{toxic} and \emph{non-toxic} reply, respectively; while 6,703 (34\%) conversations did not have a reply to the first toxic tweet. 
This suggests that toxic content may discourage others from joining or contributing to Twitter conversations. However, we see that in about half of conversations, most toxic tweets responded to non-toxic replies. We find that only a small number of conversations involve toxic initial replies to the first toxic tweets. 

\subsubsection{\textbf{Toxicity of first reply and conversation toxicity}} 
\begin{figure}
\centering
\subfloat[Toxicity distribution] {\includegraphics[width=0.45\columnwidth]{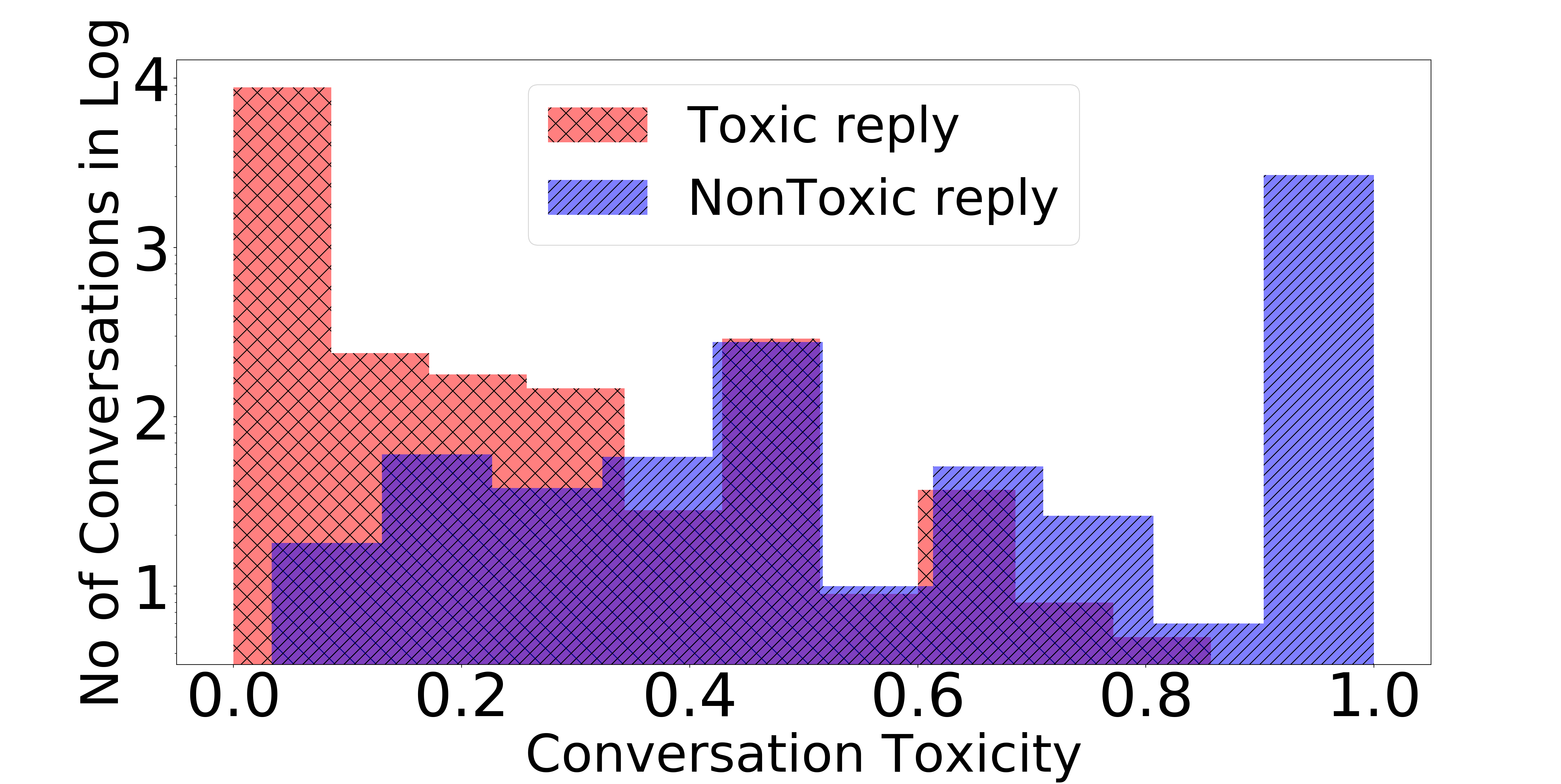} \label{fig:15}}
\subfloat[User participation]{\includegraphics[width=0.45\columnwidth]{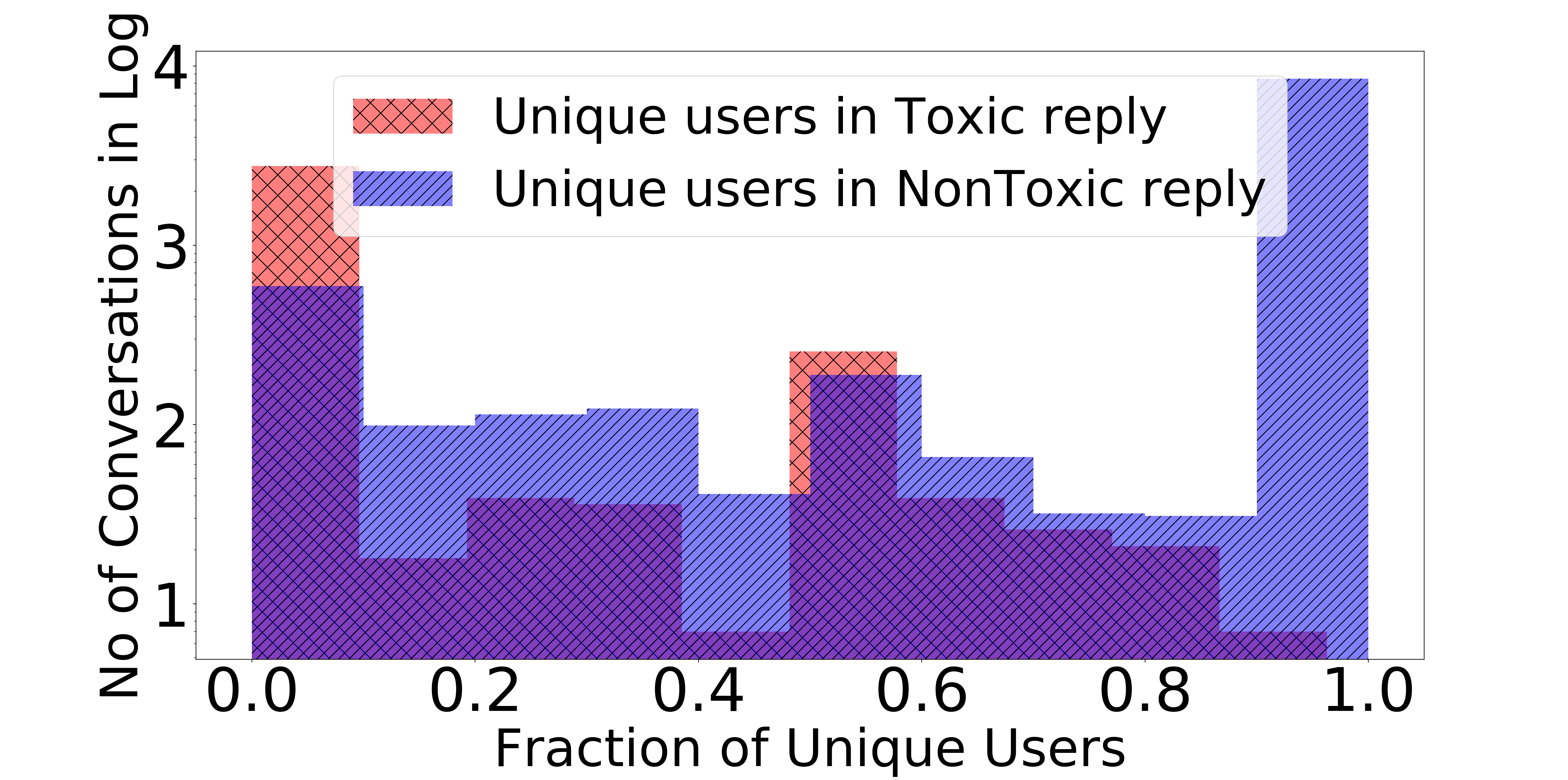} \label{fig:13}} 
\caption{Toxicity and user participation of conversations after a toxic and non-toxic reply to the first toxic tweet.}
\end{figure}

Toxic replies may lead to more toxicity within online conversations, with one hateful or toxic tweet instigating another. 
Previous scholarship has suggested as much, with hateful content spreading farther, wider and faster on social media platforms, such as Gab (Gab.com)~\cite{mathew2019spread}. 
We examine the effect of the first reply to the first toxic tweet and the toxicity of the rest of replies in the conversation.  
Figure~\ref{fig:15} shows the toxicity distribution of rest of replies in the conversations after the first toxic reply, where the red bars shows the toxicity distribution of conversations that have an immediate \emph{toxic} reply after the first toxic tweet, and the blue bars shows the toxicity distribution of conversations that have an immediate \emph{non-toxic} reply after the first toxic tweet. 
We find that conversations with a toxic reply are, on average more toxic, than conversations in which the initial reply was non-toxic (mean toxicity score of conversations with an immediate toxic and non-toxic reply is $M=0.91$ and $M=0.03$, respectively).

To examine the effect of a toxic reply to an initial toxic tweet, we construct a binary variable that is coded 0 if the first reply to the toxic tweet is non-toxic, and 1 if the first reply is toxic. We then include this toxic reply variable as a predictor in a regression analysis for the overall conversation toxicity. For this analysis, we only use the conversations that have a reply to the toxic tweet ($n=12,999$). We also employ the control variables used in previous models. 
Table~\ref{lg-remaining-toxicity} shows that if the first reply to a toxic tweet is also toxic then we expect a 0.87 increase in the predicted conversation toxicity ($p<0.001$). Thus, we find support for \emph{Hypothesis~4}: that if the initial tweet is toxic then the conversation is more likely to be toxic.  
This finding shows that toxicity may be contagious with one toxic tweet and toxic reply impacting the toxicity of conversations overall.

\begin{table} 
\centering 
\begin{minipage}{0.48\columnwidth}
\caption{Relationship between the first reply to the first toxic tweet and conversation toxicity.} 
\label{lg-remaining-toxicity}
  \resizebox{0.91\columnwidth}{!}{%
\begin{tabular}{@{\extracolsep{5pt}}lll} 
\\[-1.8ex]\hline 
\hline \\[-1.8ex] 
 \multicolumn{3}{c}{\textit{Dependent variable: Toxicity of rest of conversation}} \\ 
\cline{2-2} 
 & Linear Regression & Std err\\ 
\hline \\[-1.8ex] 
 Toxicity of first reply  & 0.87$^{***}$  & (0.003) \\ 
 to the first toxic tweet & \\
  Follower count & 0.00  & (0.00) \\ 
  Friend count & 0.00$^{*}$   & (0.00) \\ 
  Listed count & $-$0.00$^{***}$ & (0.00) \\ 
  Avg. Followers count & 0.00  & (0.00) \\
  Avg. Listed count & 0.00  & (0.00001) \\  
  Verified account & $-$0.023$^{***}$  & (0.008) \\
  Account age (years) & 0.0003  & (0.0004) \\
  Has profile image & $-$0.007$^{**}$   & (0.003) \\ 
  Has a URL & $-$0.001   & (0.003) \\
  Description length (Chars) & $-$0.00   & (0.00003) \\  
 \hline \\[-1.8ex] 
Observations & 12,999 \\ 
Log Likelihood & 6,678.316 \\ 
Akaike Inf. Crit. & $-$13,332.630 \\
\hline 
\hline \\[-1.8ex] 
\textit{Note:}  & \multicolumn{2}{r}{$^{**}$p$<$0.05; $^{***}$p$<$0.001} \\ 
\end{tabular} 
}
\end{minipage}\hfill
\begin{minipage}{0.48\columnwidth}
\caption{Relationship between the first reply to the first toxic tweet and User Participation.} 
  \label{lg-user-participation-remaining} 
 \resizebox{0.91\columnwidth}{!}{%
\begin{tabular}{@{\extracolsep{5pt}}lll} 
\\[-1.8ex]\hline 
\hline \\[-1.8ex] 
 \multicolumn{3}{c}{\textit{Dependent variable: User participation}} \\ 
\cline{2-2} 
 & Linear Regression & Std err\\ 
\hline \\[-1.8ex] 
 Toxicity of first reply & $-$0.82$^{***}$  & (0.005) \\ 
  Follower count & $-$0.00$^{***}$ & (0.00)  \\ 
  Friend count & $-$0.00$^{***}$  & (0.00)  \\ 
  Listed count & 0.00$^{**}$ & (0.00)  \\ 
  Avg. Followers count & 0.00 & (0.00) \\ 
  Avg. Listed count & 0.00003$^{**}$ & (0.00002) \\ 
  Verified account & $-$0.178$^{***}$ & (0.015) \\ 
  Account age (years) & 0.0003 & (0.001)  \\ 
  Has profile image & 0.007 & (0.006) \\ 
  Has a URL & $-$0.012$^{**}$ & (0.005)\\ 
  Description length (Chars) & $-$0.0001$^{***}$ & (0.00004) \\ 
 \hline \\[-1.8ex] 
Observations & 12,999 \\ 
Log Likelihood & $-$551.679 \\ 
Akaike Inf. Crit. & 1,127.358 \\ 
\hline 
\hline \\[-1.8ex] 
\textit{Note:}  & \multicolumn{2}{r}{$^{***}$p$<$0.001} \\ 
\end{tabular} 
}
\end{minipage}
\end{table}

\subsubsection{\textbf{Toxicity of first reply and user participation}}
We also investigate the effect of the first reply to the initial toxic tweet in conversations on user participation. This allows us to examine how replies to the first toxic tweet may influence 
how many users participation in the conversation. We define user participation as the number of unique users participating in the rest of conversation after the first toxic tweet, divided by conversation length. 
Figure~\ref{fig:13} shows user participation in conversations after the first toxic tweet. As you can see, on average the number of unique users participating after a \emph{non-toxic} reply is greater than the number of Twitter users participating in conversations after a \emph{toxic} reply.   

To examine the effect of a toxic direct reply to an initial toxic tweet on the user participation, we construct a binary variable that is coded 0 if the first direct reply to the toxic tweet is non-toxic, and 1 if the first reply is toxic. We then include this variable as a predictor in a regression analysis for the overall conversation user participation. 
For this analysis, conversations are restricted to those that have a reply to a toxic tweet ($n=12,999$). We include the same control variables as outlined previously. Table~\ref{lg-user-participation-remaining} shows that if the first direct reply to the first toxic tweet is also toxic, then we expect a 0.82 decrease in the predicted user participation ($p<0.001$). Therefore, we find support for \emph{Hypothesis~5}: that if the reply to an initial toxic tweet is toxic then the conversation is more likely be toxic. Our results suggest that how Twitter users respond to initial toxic tweets carries important consequences for the tone of subsequent direct replies.

\section{Limitations and Future Work} 
We only obtained direct replies posted within a 48 hours time span after the main tweet was posted. 
We could have potentially missed some replies, however, replies to tweets are often fast paced with most replies happening within an hour of the original post~\cite{paul2011twitter}. 
Most Twitter conversations are short with only a small minority of conversations growing to include thousands of tweets or users~\cite{goncalves2011validation}. 
Additionally, we restricted our discussion of conversations to direct replies to the main tweet. While replies to replies is another critical piece of the puzzle to understand the dynamics of toxic conversations, our goal for this study was to understand the more immediate and firsthand dynamics of toxicity within conversational exchanges on Twitter. 
Another limitation is that Twitter conversations triggered after a current event may capture different community norms of toxicity. 
We were also unable to discern the nature of toxic replies. The first reply may be a counterargument to the main tweet or a user could be replying in agreement to the main tweet using the same toxic language. Our continuous toxicity measure allows us to determine the toxicity of conversations overall but not the sentiment of tweets. 

\textbf{Future Work} 
Despite these limitations, our findings lay the groundwork for outlining strategies to decrease the toxicity online. 
Others can fully examine nested conversations to study why and when toxic exchanges spike. 
While our findings speak to differences in the Twitter accounts that participate in toxic and non-toxic conversations, 
future inquiry can build upon our work. Other work could investigate which users respond to toxic tweets and the extent to which certain users stymie or incite the spread of toxicity. 

\section{Conclusion}
Social media platforms overcome geographic distance to create a space where individuals can express opinions and engage in interpersonal communication~\cite{walther2011theories}. But with the good that online platforms offer, also comes the bad. 
To analyze the impact of toxicity on Twitter conversations, we create two metrics for toxicity. This allows us to assess the relationship between toxicity and conversational engagement, measured through the length of conversations and overall user participation in Twitter conversations. 
One key finding is that toxic conversations, defined as those with toxic tweets are longer. Another key finding is that within toxic conversations, toxicity is positively associated with user participation. 
Our findings also indicate that Twitter accounts with less identifiability are more likely to post toxic tweets. 
Twitter accounts that provide a profile image, URL, or have a verified account status were significantly less likely to participate in toxic conversations. 
Our work also highlights the importance of understanding the sequencing of toxic tweets and its impact on immediate direct replies.
 
We find that the initial direct reply to the first toxic tweet within conversations matters. Toxic replies lead to more toxic conversations. However, this suggests that the inverse is also true. Thus, non-toxic replies may potentially be the antidote to toxicity online. These findings provide key insights for developing tools to alleviate the negative outcomes of toxic content in online communities.


\end{document}